\documentclass[twocolumn,english,prl,showpacs,aps]{revtex4-2}

\usepackage[T1]{fontenc}
\usepackage[latin9]{inputenc}
\usepackage{textcomp}
\usepackage{amsmath}
\usepackage{amssymb}
\usepackage{graphicx}
\usepackage{esint}

\makeatletter
\@ifundefined{textcolor}{}
{%
 \definecolor{BLACK}{gray}{0}
 \definecolor{WHITE}{gray}{1}
 \definecolor{RED}{rgb}{1,0,0}
 \definecolor{GREEN}{rgb}{0,1,0}
 \definecolor{BLUE}{rgb}{0,0,1}
 \definecolor{CYAN}{cmyk}{1,0,0,0}
 \definecolor{MAGENTA}{cmyk}{0,1,0,0}
 \definecolor{YELLOW}{cmyk}{0,0,1,0}
}

\makeatother

\usepackage{babel}

\begin{document}

\title{Stable high-transformer ratio beam-wakefield acceleration in cusp
plasma channels }

\author{Alexander Pukhov, Lars Reichwein}

\affiliation{Institute for Theoretical Physics I, Heinrich Heine University D\"usseldorf, 40225
Germany}
\begin{abstract}
Wakefield excitation by structured electron bunches in hollow gaps
between plasma wedges is studied using three-dimensional particle-in-cell
simulations. The main part of the electron bunch has a triangular
current distribution in the longitudinal direction with a smooth head
and short tail. These bunches propagate stably in the hollow gap while
being attached to cusps of the plasma wedges. The excited wakefield
profile may have a very high transformer ratio and allows to accelerate
witness bunches to energies much higher than that of the driver bunch.
Unlike round hollow channels, where asymmetric wakefields are difficult
to avoid, no deleterious transverse beam break-up (BBU) is observed
in the gap between cusp-shaped plasma layers. 
\end{abstract}


\maketitle
Plasma-based particle acceleration is able to provide huge accelerating
fields on the order of $100\,{\rm GV/m}$ or potentially even TV/m
\cite{Leemans,Adli,Corde,Gessner,FACET,FlashFORWARD}. The wakefields
are generated either by intense laser pulses \cite{Leemans,10GeV},
or by bunches of charged particles \cite{FACET,FlashFORWARD}. Although
the laser technology is developing very fast, the currently available
particle beam drivers can provide much higher average powers. The
major problem with the beam-driven plasma wakefields is the low transformer
ratio $T_{r}=E_{a}/E_{d}$, where $E_{a}$ is the accelerating field
acting on the witness and $E_{d}$ is the drag field that is stopping
the driver. One can easily show that for a symmetric driver, $T_{r}<2$
\cite{TR2}. A beam loading of the wake by the witness will further
decrease this ratio. Thus, many stages are required to accelerate
the witness to energies much higher than that of the driver \cite{HALFH}.
An efficient coupling between multiple plasma acceleration sections
is still an unresolved problem. Thus, it important to reduce the number
of stages in a plasma-based accelerator. In an optimal case, the whole
acceleration process must be accomplished within one single plasma
stage. For the beam-driven acceleration this requires $T_{r}\gg1$,
unless the driver has already a very high energy \cite{Caldwell}.

It is well known that structured bunches can generate wakefields with
high transformer ratios. This can be continuous bunches with triangular
current profiles \cite{Bane} or tailored sequences of square bunches
\cite{Train,John}. Long drivers cannot be used in uniform plasmas
as they are subject to the self-modulation instability \cite{SMI}.
At the same time, generation of skewed wakefields and the resulting
beam break-up (BBU) \cite{BBU} limits acceleration gradients in hollow
channels \cite{Baturin}. The stable acceleration is then defined
by focusing strength of available quadrupole magnets. 

Another significant challenge is the efficient plasma-based acceleration
of positrons. While plasma allows for highly energy efficient acceleration
of electron bunches in the bubble regime \cite{Bubble,Anton}, positrons
can be accelerated in the quasi-linear regime \cite{Corde}. The latter
is rather inefficient energetically and it is difficult to conserve
emittance of positively charged witness bunches. A few exotic schemes
for positron acceleration have been proposed recently \cite{Positrons,Diederichs,Lars,Vieira,AsymHollowChan}.
These still have to be verified experimentally.

\begin{figure}
\includegraphics[width=1\columnwidth]{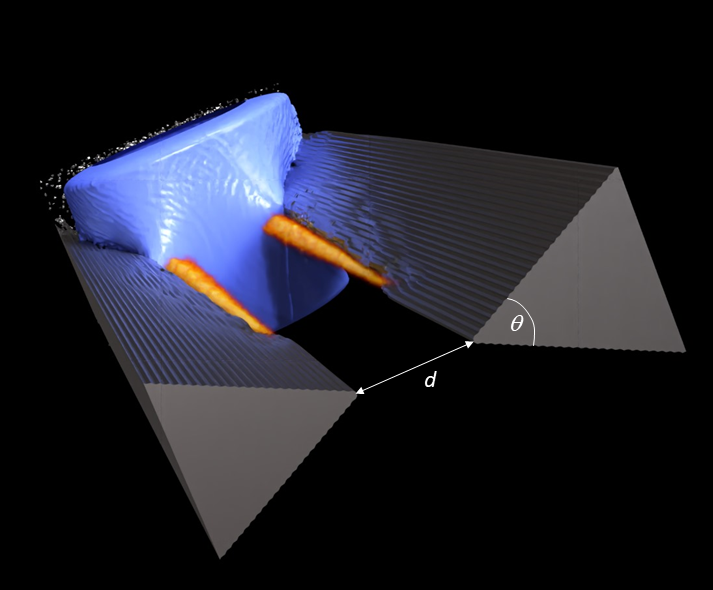}

\caption{Three-dimensional view of the simulation setup. Two driver bunches
(in yellow) are propagating stably attached to cusps of plasma wedges
and excite accelerating wakefield (in blue) with high transformer
ratio. \label{fig:wedges}}
\end{figure}

In this work, we propose a new plasma structure that naturally allows
for stable propagation of very long driver bunches and generaton of
wakefields with high transformer ratios. The plasma structure consists
of two plasma wedges with a hollow gap between the cusps as shown
in Fig.\ref{fig:wedges}. The gap width $d$ is large in comparison
with the plasma skin length, $k_{p}d\gg1$, where $k_{p}=c/\omega_{p}$,
$\omega_{p}=\sqrt{4\pi n_{e}e^{2}/m}$ and $n_{e}$ is the plasma
electron density, $e$ is the elementary charge, $m$ is the electron
density. The driver propagating inside the hollow gap has two stable
positions at the cusps of the both plasma layers. A negatively charged
driver expels plasma electrons and leaves naked ions at the edges
of the plasma layers. These ions are heavy, move rather slowly and
their focusing field attracts the driver bunch to the inner edges
of the plasma layers. 

Analytical theory of wakefields fields inside the hollow gap of such
shape is a complicated task and will be done in a separate publication.
Here, we present simulation results using the fully electromagnetic
three-dimensional (3D) particle-in-cell (PIC) code VLPL \cite{VLPL99,VLPL,RIP}.

The simulated configuration has the following parameters. Electron
density in the plasma is $n_{e}=10^{17}{\rm cm^{-3}}$. This corresponds
to the plasma skin length $k_{p}^{-1}=c/\omega_{p}\approx5.3\mu{\rm m}$.
The opening angle of the plasma wedges is $\theta=\pi/3$. The gap
width $d=3k_{p}^{-1}.$ 

The simulation box size $X\times Y\times Z$ is $100{\rm \mu m}\times100{\rm \mu m}\times430{\rm \mu m}$
with grid cells $1.5{\rm \mu m}\times1.5{\rm \mu m}\times1{\rm \mu m}$.
The simulated propagation distance was $L_{acc}=500\,{\rm cm}.$ 

\begin{figure}
\includegraphics[width=1\columnwidth]{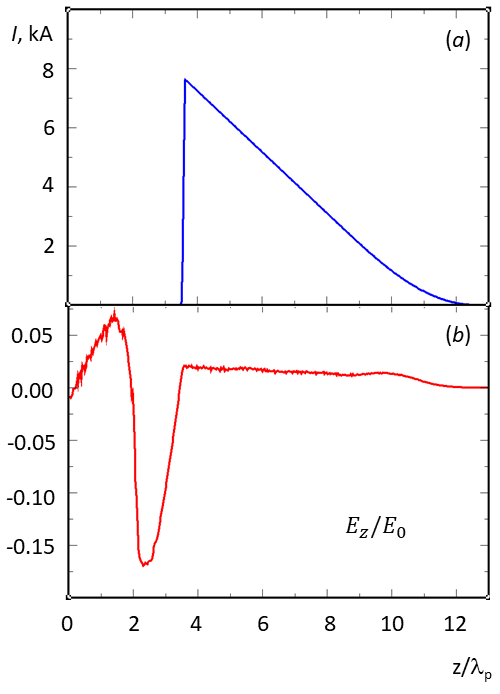}

\caption{Longitudinal profiles of the driver bunch current ($a$) and excited
wakefield ($b)$. The wakefield amplitude is normalized to the wakebreaking
amplitude $E_{{\rm 0}}=mc\omega_{p}/e$. The transformer ratio is
$T_{r}=E_{a}/E_{d}\approx7$ for this configuration. Here $E_{a}$
is the accelerating field acting on the witness and $E_{d}$ is the
drag field that is stopping the driver. \label{fig:1D}}
\end{figure}

We use two driver electron bunches with population of $1\,{\rm nC}$
each, initial energy of $10\,{\rm GeV}$ and transverse emittance
$\epsilon_{x,y}=1\,{\rm \mu m\cdot rad.}$ The driver longitudinal
current profile is triangular with a smooth leading head as shown
in Fig.\ref{fig:1D}a. The driver bunches propagate along the plasma
wedge cusps. The generated longitudinal wakefield in the center of
the gap is shown in Fig.\ref{fig:1D}b. The theoretical transformer
ratio here is $E_{\max}^{acc}/E_{\max}^{dec}\approx7$, where $E_{\max}^{acc}$
is the maximum accelerating field of the wake and $E_{\max}^{dec}$
is the maximum decelerating field acting on the driver. 

To prove the acceleration, we inject a witness bunch of test electrons
with initial energy of $10\,{\rm GeV}$ into the negative wakefield bucket
just behind the driver. The witness electron bunch has a Gaussian
density shape $n_{we}=n_{0e}\exp\left(-z^{2}/2\sigma_{ze}^{2}-r^{2}/2\sigma_{re}^{2}\right)$
with $\sigma_{ze}=7\,{\rm \mu m}$ and $\sigma_{re}=5\,{\rm \mu m}$. 

In addition, we inject a witness bunch of test positrons with the
same initial energy $10\,{\rm GeV}$ and density profile into the
positive wakefield bucket. 

\begin{figure}
\includegraphics[width=1\columnwidth]{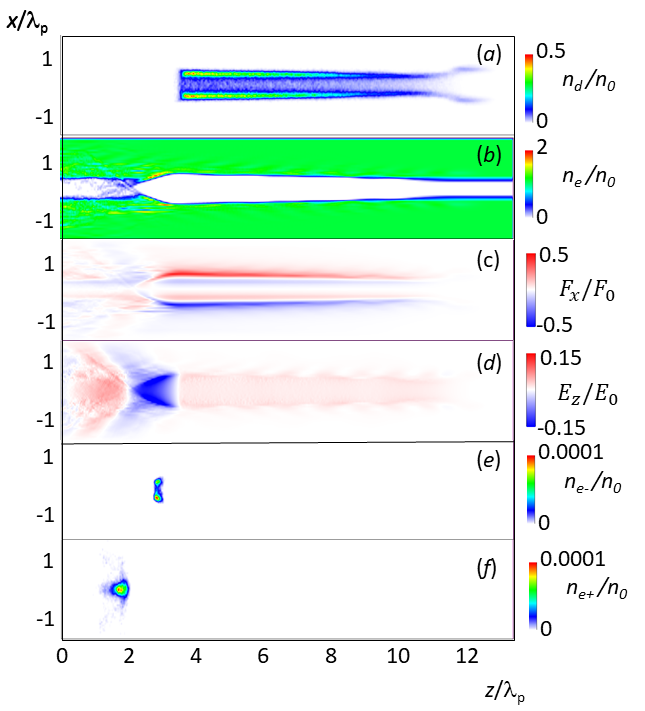}

\caption{Two-dimensional $(Z,X)$ cuts of ($a$) driver density; $(b)$ plasma
electron density; $(c)$ transverse force acting on positively charged
relativistic particles; $(d)$ longitudinal wakefield; $(e)$ density
of electron witness bunch; $(f)$ density of positron witness bunch.
\label{fig:2Dat2m}}
\end{figure}

Two-dimensional $\left(Z,X\right)$ cuts of the driver bunch density
$n_{d},$ plasma electron density $n_{e}$, $x-$component of the
radial force $F_{x}=e\left(E_{x}-B_{y}\right)$ acting on positively
charged relativistic particles running in $z-$direction, wakefield
$E_{z}$, witness electron bunch, and witness positron bunch are shown
in Fig.\ref{fig:2Dat2m} after propagation distance of $2\,{\rm m}.$
All the forces are normalized to the limiting ``wavebreaking force''
$F_{0}=mc\omega_{p}$.

The driver bunch takes on a typical ``claw''-like shape. The low
density head of the bunch propagates inside the plasma wedge. However,
as the driver density rises, its transverse electric field becomes
strong enough to expel all electrons from the plasma cusp. As a result,
the main body of the driver is attached to the cusp and remains stable
over the whole acceleration distance.

\begin{figure}
\includegraphics[width=1\columnwidth]{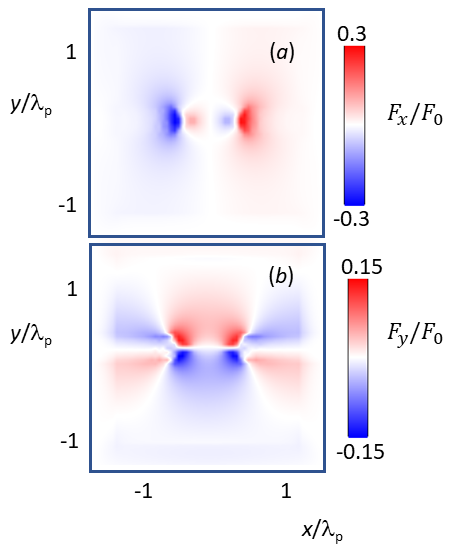}

\caption{Two-dimensional $(X,Y)$ cuts of ($a$) transverse force acting on
positively charged relativistic particles in $X-$direction; $(b)$
transverse force acting on positively charged relativistic particles
in $Y-$direction. Electrons are attracted to the cusps of plasma
wedges in the $X-$direction and to middle of the gap in the $Y-$direction.
\label{fig:4focus}}
\end{figure}

The force ${\bf F}_r$ focuses driver electrons to the cusps. Its
components $F_{x}$ and $F_{y}$ in the transverse plane $\left(X,Y\right)$
are shown in Fig.\ref{fig:4focus}. The $F_{y}$-component focuses
the driver everywhere in the gap towards the $\left(X,Z\right)$ plane.
The $F_{x}$-component defocuses the electrons from the $Z-$axis
and attracts it to the cusps positions. Inside plasma wedges, the
focusing is decaying withing the plasma skin depth. The focusing to
the cusps remains in the first negative bucket behind the driver.
Here we inject the electron witness bunch. This bunch splits into
two bunches, each being attracted to the plasma cusps. Just behind
the first negative wakefield bucket, a flash of plasma electrons is
ejected from the plasma wedge into the gap. As these electrons reach
the $Z-$axis, they focus the witness positron bunch to the axis within
the positive wakefield bucket.

\begin{figure}
\includegraphics[width=1\columnwidth]{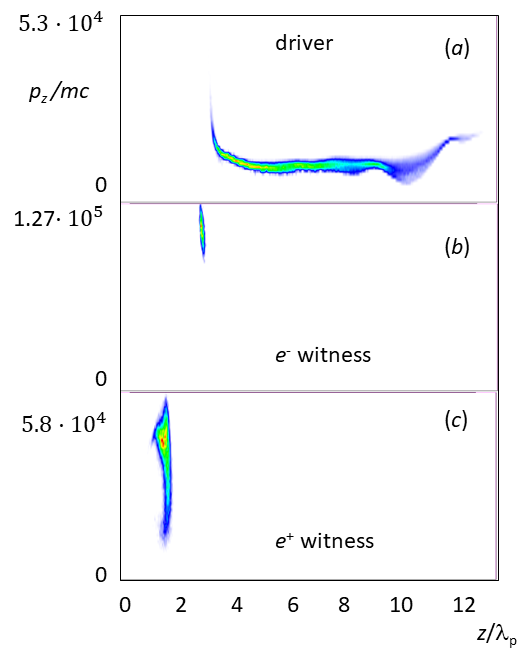}

\caption{Longitudinal phase space $(Z,P_{z})$ of $(a)$ driver, $(b)$ witness
electron bunch and $(c)$ witness positron bunch. \label{fig:6Px}}
\end{figure}

Longitudinal phase spaces of the bunches after $L_{acc}=5\,{\rm m}$
are shown in Fig.\ref{fig:6Px}. In the driver bunch, Fig.\ref{fig:6Px}a,
electrons close to the bunch head have lost up to $7\,{\rm GeV}$
energy, while the central bulk of the driver has lost about $5\,{\rm GeV}$
energy. This means, the shape of the driver does not perfectly fit
as it excites a nonlinear wake. Apparently, the transition region
between the low density ``claws'' at the head of the driver and
its bulk body excites a wakefield. Thus, the driver shape requires
further optimization. However, optimization is not a goal of the present
work and will be done elsewhere. 

\begin{figure}
\includegraphics[width=1\columnwidth]{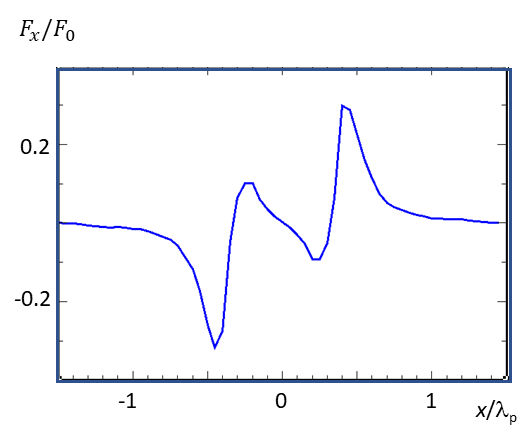}

\caption{One-dimensional $(X,Y)$ cut of transverse force acting on positively
charged relativistic particles in $X-$direction. \label{fig:5focus1D}}
\end{figure}

Electrons in the witness bunch, Fig.\ref{fig:6Px}b, have been accelerated
up to $60\,{\rm GeV}$. This gain of $50\,{\rm GeV}$ corresponds
to a transformer ratio for the electron acceleration is $T_{r}\approx7$.
It remained stable over the full acceleration distance.

The positron witness bunch, Fig.\ref{fig:6Px}c, also remained stable
and was accelerated up to $28\,{\rm GeV.}$ The energy gain of $18\,{\rm GeV}$
gives the transformer ratio for the positron acceleration about $T_{r}\approx2.5$.
It is smaller than for the electrons, because a significant part of
the wakefield energy has been lost in the transverse wavebreaking
process as the plasma electrons were sucked into the gap.

\begin{figure}
\includegraphics[width=1\columnwidth]{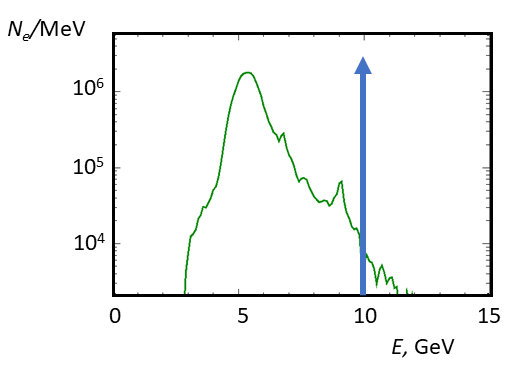}

\caption{Energy spectrum of the driver after $L_{acc}=5\,{\rm m}$ (green line)
and the initial monoenergetic spectrum at $10\,{\rm GeV}$ (blue arrow).
\label{fig:7Spectra}}
\end{figure}

\begin{figure}
\includegraphics[width=1\columnwidth]{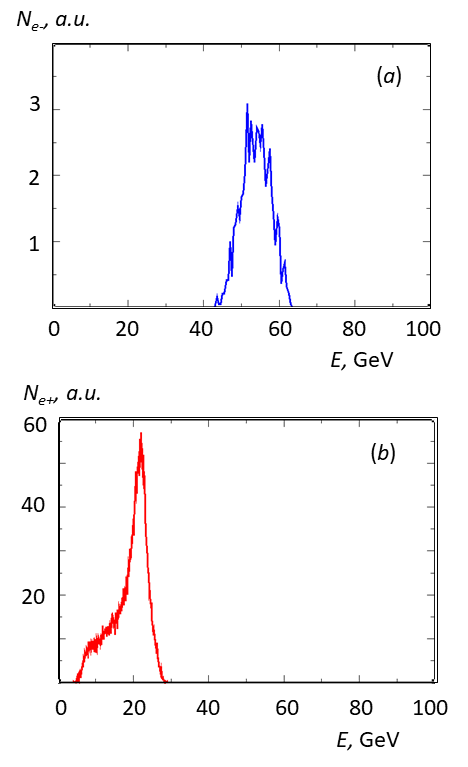}

\caption{Energy spectrum after $L_{acc}=5\,{\rm m}$ of $(a)$ the witness
electron bunch and $(b)$ the witness positron bunch. The electron
bunch acceleration corresponds to the transformer ratio of $T_{r}=E_{a}/E_{d}>7$,
while positrons show $T_{r}\approx2.5$ \label{fig:8SpectraWitnesses}}
\end{figure}

The final energy spectrum of the driver is given in Fig.\ref{fig:7Spectra}.
The main body of the driver lost energy down to $5\,{\rm GeV},$while
a few electrons in the very tail we accelerated. Spectra of witness
bunches are shown in Fig.\ref{fig:8SpectraWitnesses}a,b. These are
not monoenergetic, because we used test particles to prove the stability
of acceleration. A proper beam-loading using bunches with matched
profiles should lead to monoenergetic spectra. This optimization work,
however, is again outside of the scope of the present work and will
be done elsewhere.

Concerning the final emittance of the witness bunches, we do not provide
any numbers here as they are unreliable due to the limited numerical
resolution. We do observe a significant transverse heating of the
driver and the witnesses. However, this heating is numerical and is
reduced when we increase numerical resolution. Unfortunately, doubling
the resolution in all dimensions increases simulation time by factor
$16$, when the fully electromagnetic 3D PIC code is used. More efficient
numerical tools for simulation of this configuration have to be developed.

In conclusion, we have shown that structures consisting of plasma
wedges with a vacuum gap in between allow for stable propagation of
long electron driver bunches. Properly shaped driver excite wakefields
with high transformer ratios $T_{r}\gg2$ and can accelrate both electrons
to energies higher than that of the driver. Also positrons can be
accelerated in these structures. This may help to reduce or even avoid
staging in plasma-based acceleration and open paths towards single-stage
high energy accelerators.

This work has been supported by the Deutsche Forschungsgemeinschaft.
The authors gratefully acknowledge the Gauss Centre for Supercomputing
e.V. (www.gauss-centre.eu) for funding this project by providing computing
time through the John von Neumann Institute for Computing (NIC) on
the GCS Supercomputer JUWELS at Jülich Supercomputing Centre (JSC).

\end{document}